Comment on:
Protecting Life in the Milky Way: Metals Keep the GRBs Away
  by Stanek et al.

Adrian L. Melott[1]

ABSTRACT
Stanek et al. (astro-ph/0604113) have noted that the four low-redshift long-duration gamma-ray bursts (LSB) observed to date all occurred in faint, metal-poor galaxies. Given this selection, they argue that it is improbable that there has been a substantial population of Milky Way galaxy bursts sufficiently recently to affect life on Earth. This argument ignores the heterogeneity of stellar populations in the Milky Way, with evidence for continuing mergers with low-metallicity dwarf galaxies; observational analysis that points to LSBs being hosted by such galaxies undergoing interaction; and the existence of a likely recent GRB remnant in our galaxy.

1. Department of Physics and Astronomy, University of Kansas.  melott@ku.edu

Recently, there have been arguments that LSBs take place in low metallicity environments (Langer & Norman 2006), with the implication that the rate in our Galaxy at cosmologically recent times (e.g. Guetta & Piran 2005; hereafter GP) might be lower (Stanek et al. 2006; hereafter SGBGJKMPPW) than previously estimated. This would lower the probability of a "nearby" LSB, with implications for possible effects on terrestrial evolution, including mass extinctions (Scalo and Wheeler 2002; Melott et al. 2004; Thomas et al. 2005).

Setting aside the question of statistical significance of general conclusions based on four events, the situation is still far from clear. Environmental biases in locating GRB afterglows are poorly understood

However, Atoyan et al. (2006) have made a strong case that observation of TeV emission predicted for GRB remnants (Ioka et al. 2004) shows at least one GRB remnant in our galaxy not more than a few time $10^4$ y old. The existence of at least one such remnant is consistent with expectations based on extrapolation from the GP rate. There is some uncertainty—more remnants should be visible locally, the larger the beaming factor assumed. If SGBGJKMPPW are correct, then the existence of such a remnant is strongly disfavored for *any* reasonable beaming factor.

It is possible to make some progress by looking at processes in and around our own Galaxy, provisionally accepting the SGBGJKMPPW hypothesis on metallicity? The SGBGJKMPPW argument depends strongly upon the observed metallicity of the LSB host galaxies. The four nearby events (z<0.17) all were hosted by low-metallicity galaxies, inconsistent with the Milky Way, but possibly similar in some ways to the Small Magellanic Cloud. Thus, they conclude that an event near enough to have affected the Earth is unlikely.

The Phanerozoic Earth, ~0.5 Gy back, which has a reasonably complete fossil record, was most vulnerable to the incident radiation from a galactic LSB, since life has been accommodated to the protection from solar UVB afforded by the ozone shield, which is strongly depleted by LSBs within a few kpc (Thomas et al. 2005). Therefore it is interesting to consider our galaxy's star formation history on that timescale.

Although the LMC and SMC are unlikely to have had more than one encounter with the extended disk of the Milky Way during the Phanerozoic, our galaxy appears to have a history of merger events with low-metallicity satellites with characteristics not inconsistent with those inferred by SGBGJKMPPW for LSB host galaxies. There are many examples and indications of continuing merger activity. For example, the Sagittarius dwarf galaxy is in a close orbit with period 0.85-1 Gy, and apogalactica 12-15 kpc (Ibata et al. 1997; Law et al. 2005). Studies of stellar populations in the bulge indicate young (~200 My) low metallicity stars, [M/H]~-1.5 to -2 (van Loon et al. 2003). This population contains stars of lower metallicity than older stars in the same region. Metal-poor gas to fuel such star formation must originate outside the central regions of the galaxy. One suggestion is the infall of metal-poor gas from the galactic halo such as the High-Velocity Clouds (Richter et al. 2001); of course gas-rich dwarf galaxies will do as well.  Another very recent merger event is indicated for the outer disk in a study of Cepheids (Yong et al. 2006).  Again, the young stars are of lower metallicity than the older clusters and field stars.  Taken together, these suggest accretion events by low-mass, low-metallicity dwarf galaxies into the Milky Way possibly every few 100 My.  There are more.  The Monoceros Ring detected by the SDSS (Newberg et al. 2002) is probably associated with disruption of the Sagittarius Dwarf galaxy. There is a candidate Canis Major dwarf galaxy at the outer disk (Martinez-Delgado et al. 2005, and references therein).  This week saw the announcement of a new, faint, very metal-poor dwarf galaxy in Boötes at an estimated distance of 60 pc (Belokurov et al. 2006).  Dwarf irregulars are even more difficult to detect than dwarf spheroidals.

A collision between any gas-rich object and the galaxy is likely to increase the star formation rate.  Massive stars originating in the ISM of the low-metallicity dwarf are reasonable LSB progenitors.  How does this hypothesis agree with observations of LSB host galaxies?

LSB host galaxies include a high fraction of merging and interacting systems, *independent of redshift and galaxy luminosity* (Wainwright et al. 2005). The fraction of mergers is elevated compared with other high redshift samples, particularly considering the low luminosities of the GRB hosts.

Thus, the GRB host population is consistent with a picture in which many bursts arise in low metallicity, interacting galaxies, and the recent history of our galaxy is consistent with many interactions of this type.  Arguments based on the mean metallicity of the disk, without taking account of its environment, miss this likely scenario.

5. ACKNOWLEDMENTS
I thank Bruce Twarog for helpful comments This work was supported by the NASA Astrobiology: Exobiology and Evolutionary Biology Program under grant NNG04GM14G.

REFERENCES

Atoyan, A. 2006 ApJ Lett, in press (astro-ph/0509615)

Guetta, D., & Piran, T. 2005 A&A 435, 421

Ibata, R.A. et al. 1997 AJ 113, 634

Ioka, K., Kobayashi, S., & Mészaros, P. 2004 ApJ 613, L17.

Langer, N., & Norman, C. 2006 ApJ 638, L63

Law, D.R. et al. 2005 ApJ 619, 807

Martinez-Delgado, D. et al. 2005 ApJ 633, 205

Melott, A.L. et al. 2004 Int. J. Astrobiology 3, 55 (astro-ph/0309415)

Newberg, H.J. et al. 2002 ApJ 569, 245

Richter, P. et al. 2001 ApJ 559, 318

Scalo, J., & Wheeler, J.C. 2002 ApJ 566, 723

Stanek, K.Z. et al. 2006 ApJ, submitted (astro-ph/0604113)

Thomas, B.C. et al. 2005 ApJ 634, 509

van Loon, J. et al. 2003 MNRAS 338, 857

Wainwright, C. et al. 2005 ApJ, submitted (astro-ph/0508061)

Yong, D. et al. 2006 AJ 131, 2256